\documentclass[conference]{IEEEtran}
\IEEEoverridecommandlockouts
\usepackage{tabularx} 
\usepackage{cite}
\usepackage{amsmath,amssymb,amsfonts}
\usepackage{algorithmic}
\usepackage{graphicx}
\usepackage{svg}
\usepackage{textcomp}
\usepackage{xcolor}
\usepackage[most]{tcolorbox}
\usepackage{url}
\usepackage{enumitem}
\def\BibTeX{{\rm B\kern-.05em{\sc i\kern-.025em b}\kern-.08em
    T\kern-.1667em\lower.7ex\hbox{E}\kern-.125emX}}
\tcbset{textmarker/.style={
        enhanced,
        parbox=false,boxrule=0mm,boxsep=0mm,arc=0mm,
        outer arc=0mm,left=5mm,right=3mm,top=7pt,bottom=7pt,
        toptitle=1mm,bottomtitle=1mm}}

\newtcolorbox{noteBox}{textmarker,
    borderline west={4pt}{0pt}{gray},
    colback=gray!10!white}
\newcommand{\note}[1]{\begin{noteBox} #1 \end{noteBox}}
    
\begin{document}

\title{Code Collaborate: Dissecting Team Dynamics in First-Semester Programming Students}

	\author{\IEEEauthorblockN{
        \begin{minipage}{0.4\textwidth}
            \centering
            Santiago Berrezueta-Guzman\\
            \textit{Technical University of Munich}\\
            Heilbronn, Germany \\
            s.berrezueta@tum.de
        \end{minipage}
        \hspace{0.1\textwidth}
        \begin{minipage}{0.4\textwidth}
            \centering
            Patrick Bassner\\
            \textit{Technical University of Munich}\\
            Munich, Germany \\
            bassner@in.tum.de
        \end{minipage}
    }
    \vspace{1\baselineskip} 
    \IEEEauthorblockN{
        \begin{minipage}{0.4\textwidth}
            \centering
            Stefan Wagner\\
            \textit{Technical University of Munich}\\
            Heilbronn, Germany \\
            stefan.wagner@tum.de
        \end{minipage}
        \hspace{0.1\textwidth}
        \begin{minipage}{0.4\textwidth}
            \centering
            Stephan Krusche\\
            \textit{Technical University of Munich}\\
            Munich, Germany \\
            krusche@tum.de
        \end{minipage}
    }
	}

\maketitle

\begin{abstract}

Understanding collaboration patterns in introductory programming courses is essential, as teamwork is a critical skill in computer science. In professional environments, software development relies on effective teamwork, navigating diverse perspectives, and contributing to shared goals. This paper offers a comprehensive analysis of the factors influencing team efficiency and project success, providing actionable insights to enhance the effectiveness of collaborative programming education. By analyzing version control data, survey responses, and performance metrics, the study highlights the collaboration trends that emerge as first-semester students develop a 2D game project.

Results indicate that students often slightly overestimate their contributions, with more engaged individuals more likely to acknowledge mistakes. Team performance shows no significant variation based on nationality or gender composition, though teams that disbanded frequently consisted of “lone wolves,” highlighting collaboration challenges and the need for strengthened teamwork skills. Presentations closely reflected individual project contributions, with active students excelling in evaluative questioning and performing better on the final exam. Additionally, the complete absence of plagiarism underscores the effectiveness of proactive academic integrity measures, reinforcing honest collaboration in educational settings.
\end{abstract}


\section{Introduction}\label{sec:introduction}

Team and collaborative projects have emerged as pivotal elements of university education, underscoring the shift towards more engaging and practical learning methodologies \cite{oakley2004turning}. These approaches improve students' technical skills and polish essential soft skills such as communication, conflict resolution, and project management—competencies that are indispensable for professional success\cite{chorfi2022problem}. Collaborative projects offer a dynamic platform for students to apply theoretical knowledge through peer learning and collective problem-solving, mirroring real-world scenarios where interdisciplinary collaboration is crucial to enhancing the educational experience and making it more inclusive and supportive\cite{prada2022teamwork}. By fostering a culture of collaboration, universities prepare computer science students to thrive in a field where innovation is often the product of team effort \cite{yadav2021collaborative}.

This study explores key aspects of collaboration and team dynamics among first-semester programming students, focusing on areas that have yet to be extensively studied and answering important research questions (RQ). 

\begin{enumerate}[label=\textbf{RQ\arabic*}, leftmargin=*]
    \item Do teams with members from a single nationality perform differently than those from multiple nationalities?
    \item Is there a difference in performance between teams of members of the same gender and teams of members of different genders?
    \item How accurately do participants estimate their contribution to the project development?
    \item How accurately do participants estimate their errors during project development?
    \item What were the most utilized resources in the project development?
    \item Is it possible to prevent plagiarism in this kind of activity?
    \item Is there a link between a student’s project backlog and their performance in project presentations?
    \item Is there a correlation between the grades received on the project and the intermediate and final exam grades?
\end{enumerate}

Therefore, we compared team performance based on composition, explicitly examining differences between teams of the same nationality versus mixed-nationality teams and same-gender versus mixed-gender teams (RQ1 and RQ2). Additionally, we investigated the 'lone wolf' approach in team projects and its impact. We also assessed how accurately students estimate their contributions and mistakes during project development (RQ3 and RQ4). A study identified the most commonly used resources in project development (RQ5) and explored strategies to prevent plagiarism between teams (RQ6). Furthermore, we analyzed the relationship between project backlogs and presentation performance (RQ7) and looked for a correlation between project grades and individual final exam grades (RQ8). 

Answering the research questions above, the research advances the understanding of team dynamics, offering comprehensive insights to optimize collaborative learning and prepare students for professional teamwork environments. 

This paper is structured as follows: Section \ref{RW} reviews related studies, emphasizing distinctions between the research in this paper and other contributions. Section \ref{M} outlines the methodology employed in this study. We present the results in Section \ref{R} and discuss them in Section \ref{D}. Finally, Section \ref{C} summarizes the paper, highlights key findings, and outlines future research challenges.

\section{Related Work}\label{RW}

The analysis of teamwork behaviors in educational settings, particularly within software development, has garnered significant attention due to its implications for academic performance and professional preparedness. Various studies have explored different aspects of teamwork, leveraging diverse methodologies and theoretical frameworks to understand and enhance collaboration among students \cite{soundarajan2015collaborative, neill2017improving}.

Gitinabard et al.~\cite{gitinabard2020student} explored the use of GitHub logs\footnote{\textit{GitHub logs} are records of events that occur within a GitHub repository, such as commits, pull requests, issues, and other activities, allowing users to track changes, \url{https://docs.github.com/en/get-started/using-git/about-logs}.} to analyze teamwork behaviors in university programming projects, underscoring the significance of teamwork facilitated through version control systems in both educational and professional programming contexts. The research categorizes student teams into collaborative, cooperative, or solo-submit groups by analyzing contribution patterns in GitHub logs from a CS2 Java course. Key findings reveal that the division of labor and contributions within teams can be effectively measured and classified using GitHub activity logs. The teamwork style substantially impacts project outcomes, with balanced contributions correlating with enhanced performance and satisfaction. Additionally, the study suggests that these insights could lead to automated tools to support teamwork and peer learning, demonstrating the potential of version control data to assess and improve the quality of teamwork in educational programming projects. This approach offers a promising avenue to optimize educational practices and prepare students for professional environments.

Subsequently, Gitinabard et al.~\cite{gitinabard2023analysis} take a more advanced approach by applying machine learning techniques (self-supervised and semi-supervised learning) to automatically classify teamwork styles based on GitHub interactions. The research further examines how different teamwork styles, particularly balanced contributions, correlate with academic performance, providing deeper insights into how automated analysis of version control data can support effective student collaboration. The findings confirm that teams with more balanced contributions tend to achieve better academic results, though it notes that their help-seeking behavior is not significantly different from other groups. It demonstrates the effectiveness of using machine learning to classify and understand teamwork styles, highlighting the role of balanced contributions in educational success.

Sancho-Thomas et al.~\cite{sancho2009learning} examined the effectiveness of the NUCLEO e-learning framework in university programming courses, emphasizing the integration of teamwork skills essential for real-life software development projects. This research demonstrates how a socio-constructivist pedagogical approach, facilitated by problem-based learning within the NUCLEO framework, enhances student engagement and actively promotes the acquisition of crucial teamwork skills such as communication, leadership, and conflict management. The findings underline the importance of structured, technology-supported educational frameworks that simulate real-world environments, thereby improving both the learning process and outcomes in programming education.

Richards and Bilgin~\cite{richards2012cross} explored the differences in teamwork behaviors and attitudes between Information and Communication Technologies (ICT) students in Australia and Singapore within virtual global teams. Utilizing an online survey that assessed temporal and cultural dimensions of teamwork, the study highlighted significant differences in temporal behaviors such as punctuality and time management, reflecting more profound cultural influences. While the study confirmed some cultural stereotypes, it noted only minor differences in other cultural dimensions and pointed out gender-specific differences in team behavior. The findings underscore the necessity for educational providers to integrate cross-cultural training in their curricula to better prepare students for the dynamics of global teamwork, suggesting that such preparation is crucial for students entering the global workforce. This study provides insights into the impacts of cultural and temporal differences on teamwork in an educational setting, advocating for tailored educational approaches to enhance global teamwork competencies.

Iacob and Faily~\cite{iacob2019exploring} investigated the gap between undergraduate students' initial expectations and experiences following a 27-week software engineering course based on a group project. The study reveals a significant discrepancy between students' anticipations regarding teamwork dynamics and project management and real-world encounters at the course's end. It shows that students underestimate collaboration challenges, such as conflict resolution and effective coordination, leading to a stark contrast between expected and actual team functionality. The research emphasizes the necessity for educational strategies to better prepare students for realistic project scenarios, suggesting a need to align academic frameworks more closely with real-world software development teamwork complexities.

Simpson and Storer~\cite{simpson2017experimenting} explored substantial enhancements to software engineering education at the University of Glasgow to bridge the gap between academic learning and real-world application. Essential modifications included the introduction of real-world customers to provide genuine project requirements, adopting a flipped classroom model to enhance hands-on learning, emphasizing the impact of technical debt through targeted assessments, and using senior students as mentors to enrich educational experiences. These changes resulted in increased student engagement and more realistic preparation for professional software development practices, illustrating the significant benefits of incorporating realistic project experiences within software engineering education.

In contrast to previous studies, which relied exclusively on static measures like GitHub logs, the approach in this paper challenges established findings by investigating the potential impact of team composition based on nationality or gender on project outcomes. Furthermore, we analyze students' commitment levels throughout the project duration and their self-perceived contributions and errors in project development. This study also examines the resources utilized in project development and assesses the effectiveness of proactive measures in maintaining academic integrity, particularly in preventing plagiarism. Finally, we explore any correlations between collaborative work dynamics and individual performance evaluations. 

\section{Method}\label{M}

This section delineates the research method, data acquisition protocols, and analytical analysis employed in this paper. 

\subsection{Course Selection and Demographic Analysis}

The study sample consisted of 114 first-semester information engineering students enrolled in an introductory programming course at the Technical University of Munich. The course is based on interactive learning \cite{krusche2017interactive, krusche2023introduction} and includes several assessments ranging from computer-based exams to a team project.
The students were grouped into 57 teams, chosen at their discretion. This sample comprised students from 31 nationalities, and 27 \% of them were women.
Seven students took the course a second time after failing the previous year, bringing valuable perspectives from students with prior exposure to the course content, although not to the project introduced this semester. Apart from these repeat students, the entire sample had not previously encountered one another in other educational settings, adding an element of novelty to the study.
14 \% of the sample had prior experience in incomplete university programs in their countries, such as other computer science programs.

\subsection{Tools}
This course utilizes the Artemis learning platform \cite{krusche2018artemis} to create and manage lectures, exercises, and examinations \cite{linhuber2024}. Artemis includes a plagiarism detection feature powered by JPlag\footnote{JPlag is a system that finds pairs of similar programs among a given set of programs. \cite{prechelt2000jplag}}, which identifies similarities between student submissions. Additionally, the platform provides access to the students' repositories and commit histories for programming exercises stored in Bitbucket\footnote{Bitbucket is a web-based version control repository hosting service enabling teams to manage and collaborate on code projects.}. 
Therefore, we used Artemis to publish and grade the project.

\subsection{Description of the Project}
The project involves developing a well-known 2D game in the style of \textit{Maze Runner}. The game principle is guiding a character through a labyrinthine maze. The objective is to navigate from an entry point to an exit, overcoming various obstacles, including traps, adversaries, and a lock (exit) that requires a key. The maze comprises numerous interior walls, creating a complex pathway with challenging routes and dead-ends. Therefore, players must strategically move the character, avoid or confront randomly placed traps and enemies, and locate keys to unlock the exit to complete the game. 

\subsubsection{Gameplay Requirements}
Students must fulfill specific requirements to attain a perfect grade on this project.

\textbf{Walls and Paths}: Each maze features an entrance and one or more exits along its outer perimeter, encased by solid walls without openings. The entrance, which may reside anywhere within the maze or along its border, serves as the starting point for the player. The exits, on the other hand, are the ultimate goal, positioned along the outer perimeter. Clear distinction between the entrance, exits, and walls is essential for player navigation.
In the game, 'paths' denote accessible areas where players can move. Maze layouts vary significantly; some are spacious with few walls, facilitating more effortless movement, while others are densely packed with narrow passages.

\textbf{The Character} moves vertically or horizontally within the maze, restricted to open spaces and unable to pass through walls. The character begins with a set number of lives, with the game concluding if all lives are lost before successfully reaching and unlocking an exit.

\textbf{The Obstacles}: These elements can cause the character to lose a life upon contact.

\begin{itemize}
    \item \textbf{Traps}. Fixed obstacles placed along paths that result in life loss if touched.
    \item \textbf{Enemies}. Mobile obstacles that move randomly through the maze and cannot pass through walls.
\end{itemize}

\textbf{The Key}: Essential for exiting the maze, the character must collect the key to unlock the exit; otherwise, the exit remains inaccessible like a wall.

\textbf{The heads-up display (HUD)}: Displays the number of lives remaining and indicates whether the key has been acquired. 

\textbf{The Game Menu}: Immediately appears at game start and can be accessed by pressing \textit{Esc} during gameplay, pausing all in-game actions. Options include resuming the game, starting a new game with a different map, or exiting the game entirely.

\textbf{The Victory and Game Over display}: Appears upon successfully exiting the maze without losing all lives ("You won") or losing all lives before exiting ("Game Over"). In both scenarios, the game stops, allowing players to return to the game menu to retry or choose another action.

\subsubsection{Technical Requirements}

In the game, \textbf{mazes} are not created directly in the program's code. Instead, the game is designed to load any maze provided in a Java properties\footnote{Java Properties is a simple means to store and load key-value pairs from a file, see \url{https://docs.oracle.com/javase/8/docs/api/java/util/Properties.html}} file. This file format records data as key-value pairs, similar to a map where both the key and the value are strings. In the context of this game, the keys represent the coordinates of the maze cell, formatted as 'x, y'. The value assigned to each key indicates the type of object at that coordinate, with several types defined: '0' for a wall, '1' for an entry point, '2' for an exit, '3' for a static trap, '4' for a dynamic enemy, and '5' for a key.

When a player selects the 'load map' option from the game menu, a file chooser window opens, allowing them to choose a maze file. Once a file is selected, the game reads this file, interprets the maze structure based on the coordinates and objects defined, and starts the game with the maze setup from the chosen file. Players can see these configurations by checking the files in the 'maps' directory.

The game is developed as a 2D top-down game using the libGDX framework\footnote{LibGDX is a versatile, open-source Java framework that allows developers to create 2D and 3D games for multiple platforms, including Windows, macOS, Linux, Android, iOS, and web browsers \cite{stemkoski2018libgdx}.}. Each game object, identified by coordinates in a properties file, is represented by a 2D sprite.

The game includes a camera movement mechanism to keep the player character visible. During gameplay, this camera holds the player within the middle 80 \% of the screen horizontally and vertically. If the game window is resized, the camera automatically adjusts to maintain this visibility rule. The game adapts to different window sizes without scaling the game elements. Regardless of the window size, each sprite retains its original size of 16x16 pixels. A larger window allows more of the maze to be visible at once, enhancing the player's view of the game environment without altering the graphical scale of the elements.

The game incorporates music and sound effects using the \textit{Music} and \textit{Sound} classes provided by the libGDX framework. It's crucial to ensure the audio levels between music and sound effects are well-balanced so that neither is overpowering. During gameplay, a background music track loops continuously, selected to match the intensity and theme of the game to enhance the immersive experience. A different, calmer music track is played for the game menu to create a more relaxed atmosphere.

The game has to be developed using object-oriented programming principles. Each object type within the game—such as walls, entry points, exits, traps, and enemies—is represented by a dedicated class. These classes inherit from a common superclass named \textit{GameObject}, which encapsulates shared functionality and properties. This design allows for code reusability and reduces duplication through inheritance and delegation. This approach organizes the code more efficiently and simplifies modifications and maintenance.

The game's code must be thoroughly documented using JavaDoc, ensuring every class and method includes clear and informative comments. A \textit{README} file accompanies the project, outlining the overall code structure and class hierarchy to help reviewers quickly grasp the organization and function of the components. Additionally, the \textit{README} details the steps necessary to run and interact with the game and explains the game mechanics that extend beyond the basic requirements.

\subsubsection{Academic Integrity Guidelines}

While students are allowed to use any resources of their choice for their projects, they are informed that a plagiarism check will be conducted immediately after the deadline and before the project presentation. They are aware that significant penalties will be imposed on teams with substantial code similarities that cannot be justified.

We did not restrict the use of artificial intelligence (AI) tools, such as ChatGPT. Although previous research has identified techniques to avoid its use \cite{berrezueta2023recommendations}, we decided that due to the project's complexity, students can use these tools. However, we strongly encourage a deep understanding of the material over simple copy-paste implementations.

\subsubsection{Template}

Each team has a repository template to clone and use to commit to their implementation on Artemis. When students clone and run the template, an initial screen with background music and a "Start Game" button is displayed. Upon clicking this button, a character moves randomly on the screen, accompanied by a message indicating that pressing ESC will return to the previous menu.

The project description includes the minimum requirements and a troubleshooting section to help students with configuration issues based on their operating system.

\subsubsection{Grading Criteria}

The project grade contributes 30 \% to the total course grade. The remaining 70 \% was divided between two exams and the presentation of homework assignments. The project is graded out of 100 points, as detailed in Table \ref{grading}.
 
\begin{table}[!ht]
\centering
\caption{Assessment Criteria for each category in the Game Development Project}
\label{grading}
\begin{tabularx}{\columnwidth}{|p{1.6cm}|X|c|}
    \hline
    \textbf{Category} & \textbf{Criteria} & \textbf{Points} \\
    \hline
    Game \newline world & The game has one entrance and multiple exits. Reachable exit. Interesting and challenging level design with logical connections and value-added areas. Placed obstacles include static traps and dynamic enemies with unique behaviors. Keys for opening exits. & 20 \\
    \hline
    Main \newline character & Character movement in four directions. Collision system preventing movement through obstacles. Multiple lives with various ways to lose them. The camera follows the character. & 15 \\
    \hline
    Graphical User \newline Interface (GUI) & Game menu with start, settings, and exit options. Pause menu. Base HUD with important information styled to fit game design. Victory and Game Over menus with transitions to the main menu. & 10 \\
    \hline
    Sound \newline design & Background music during gameplay and menus. Sound effects for various game events. & 10 \\
    \hline
    Graphics & Appropriate visual style and rendering on different devices without discomfort or graphical issues. & 10 \\
    \hline
    Code \newline Structure & Object-oriented implementation. There is no code duplication. Proper use of inheritance, delegation, and method extraction. & 15 \\
    \hline
    Documentation & JavaDoc for most methods. Comments in long methods. README file detailing code structure, game use, and mechanics. & 10 \\
    \hline
    Soft skills & Efficient use of time for a 15-minute presentation. Clear and structured communication with a good opening and closing. & 10 \\
    \hline
\end{tabularx}
\end{table}

\subsubsection{Bonus Implementation}

In addition to fulfilling the minimum requirements, students are encouraged to incorporate additional functionalities that can earn them bonus points, potentially increasing their total grade by up to 10 \%. These bonus points are distributed across various categories in Table \ref{bonus}. It is crucial to note that bonus points are not awarded for code structure, documentation, and soft skills achievements. Furthermore, it is a prerequisite that students first meet the minimum requirements to qualify for bonus points.

\begin{table}[!ht]
    \centering
    \caption{Assessment Criteria for providing bonus points in each grading category.}
    \label{bonus}
    \begin{tabularx}{\columnwidth}{|>{\centering\arraybackslash}m{1.3cm}|X|>{\centering\arraybackslash}m{1.2cm}|}
    \hline
    \textbf{Category} & \textbf{Criteria} & \textbf{Bonus points} \\
    \hline
    Game World & For additional features and interesting obstacles & up to 3 \\
    \hline
    Main Character & For additional player mechanics and features & up to 3 \\
    \hline
    Graphics & For unique art style and high level of details & up to 2 \\
    \hline
    GUI & For unique, responsive, and user-friendly UI & up to 1 \\
    \hline
    Sound Design & For layered sounds that combine to enhance the atmosphere of the game or improve the impact of the player's abilities & up to 1 \\
    \hline
    \end{tabularx}
\end{table}

\subsection{Implemented Analysis}
\begin{figure}[h] 
	\centering 
	\includegraphics[width=1\columnwidth]{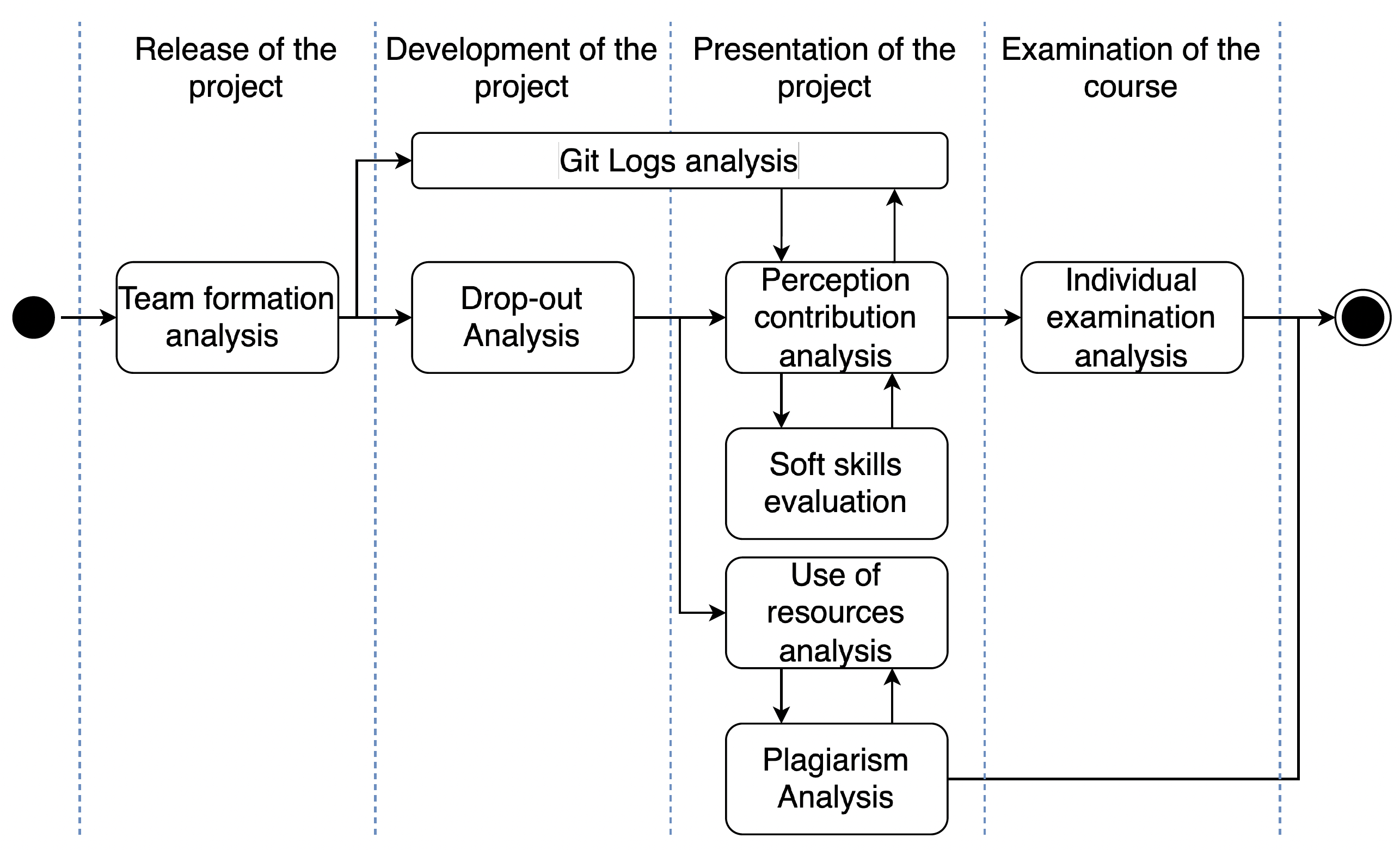} 
	\caption{Overview of the study implemented in this research across each significant period.}
	\label{analysis} 
\end{figure}

We explored several key patterns in project resolution to address the research questions outlined in this study (RQ1 -- RQ8). This analysis is structured over time, as depicted in Figure \ref{analysis}.

\subsubsection{Team Formation Analysis} We first examine the students' tendencies to form teams for collaborative projects, analyzing how these teams are formed based on nationality, language, and gender. We also analyzed patterns in dropout cases during the development of this project until its presentation.  

\subsubsection{Git Logs Analysis} We evaluated each team member's commitment level throughout the project development period by analyzing their weekly contributions to the team's repository. This involved reviewing the repository history and counting each student's significant weekly commits, including bug fixes, documentation updates, and implementation work. For this and the subsequent analyses, we assigned a number to each student in the team, where "student 1" is the author of the first commit, and "student 2" is the other team member. 

\subsubsection{Perception Contribution Analysis} Upon completion of the project presentations, all teams completed a survey that included questions regarding their perception of their contributions to each aspect of the project development (Game world, main character, GUI, sound design, graphics, documentation, and bonus points). During the survey, team members were not allowed to communicate with each other or share their responses. The results of the students' self-assessments were then compared with an estimation based on the number of significant commits made in the team's repository. This comparison allows us to analyze whether students overestimate or underestimate their contributions.

A similar analysis was conducted to assess the occurrence of errors during project development. Errors were defined as commits that resulted in build failures, merge conflicts, or those followed by refactoring due to poor-quality code or bug fixes. Students were then asked to estimate the percentage of these errors or setbacks they believed were their responsibility.

\subsubsection{Use of Resources Analysis} Each team member indicated the percentage of resources utilized to complete the project in the survey. We provided predefined options, including artificial intelligence tools, lecture slides, forums, and the dedicated Discord-server\footnote{This course uses Discord as an official communication channel among students and instructors.}. However, the students can specify additional resources they utilized for project completion. 

\subsubsection{Soft Skills Evaluation} During the project presentations, we evaluated each team member based on their proficiency in efficient, clear, and structured communication skills. We also assessed their performance in responding to questions from assessors (instructors), focusing mainly on students with the lowest commitment contributions in GitHub logs. The entire team was also evaluated on their ability to manage allocated time during the demo and to showcase any additional features implemented.

\subsubsection{Plagiarism Analysis} Before the presentation of the project, we conducted a plagiarism check to identify potential misconduct, such as sharing code between teams. 

\subsubsection{Individual Examination Analysis} We investigated correlations between the grades obtained in the project and the intermediate individual examinations of the course.

\section{Results}\label{R}

\subsection{Team Formation and Drop-out Analysis} 

\begin{figure}[h] 
 	\centering 
 	\includegraphics[width=1\columnwidth]{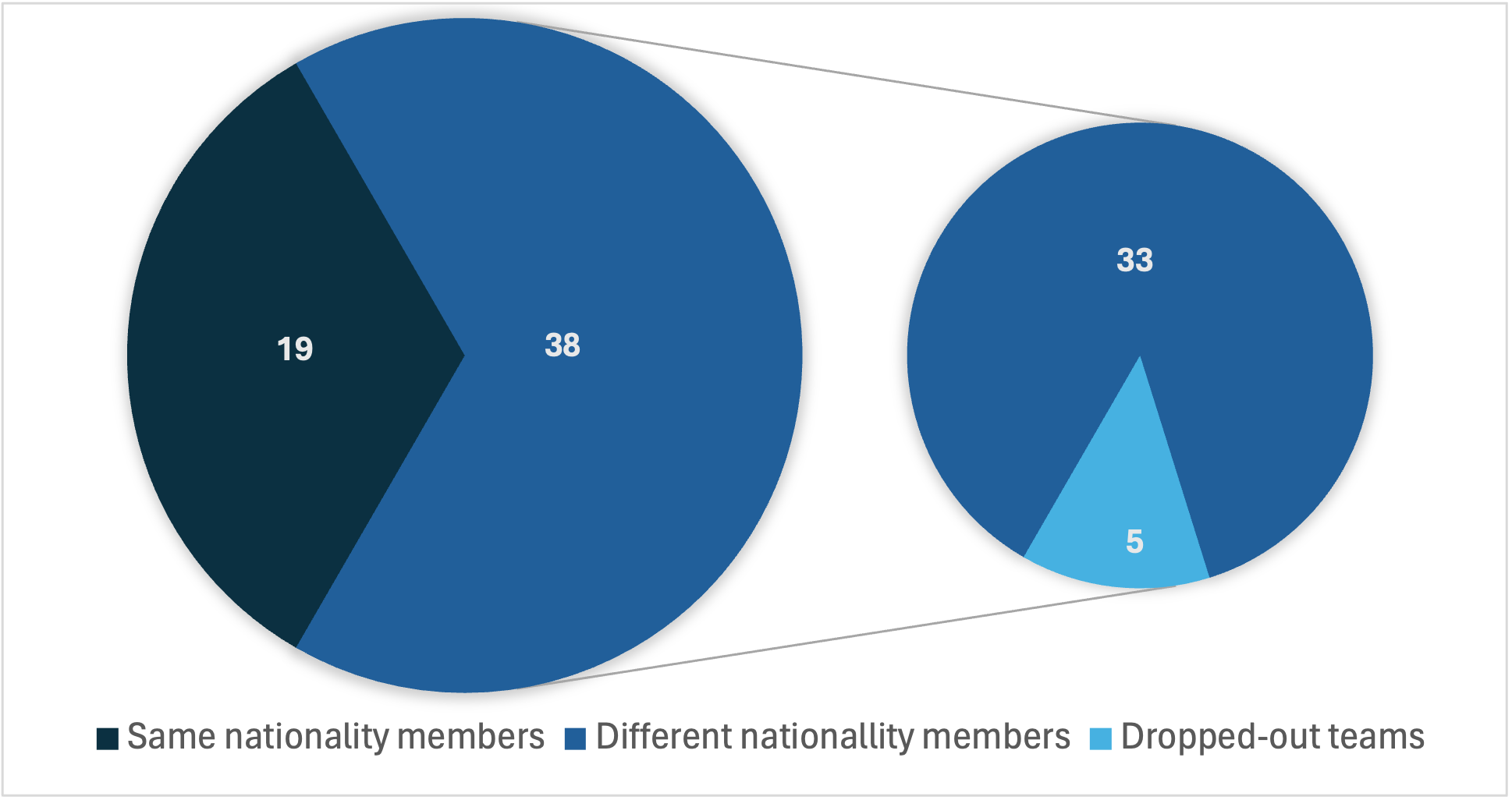} 
 	\caption{ The distribution of team formation trends indicates a notable involvement of students from diverse nationalities, particularly within teams that experienced dropout cases during the project. }
 	\label{demography} 
 \end{figure}

One initial finding presented in Figure \ref{demography} is the trend among first-year students in choosing their team partners. Within a diverse group of students from 31 nationalities, it was observed that students selected teammates of the same nationality in 33~\% of the cases; this means that 19 out of 57 teams were formed of students of the same nationality. We believe this is due to the ease of communicating in their native language, which, in this context, differs from English.

Furthermore, five of the 57 teams dropped out of the project without attempting it. These teams were composed of members from different nationalities; one of the possible causes could be that communication is more effective among individuals who share the same national background and may have had prior interactions during the course. We found no significant difference in performance between uni-nationally-formed and multi-nationally-formed teams (r = 0.08; p = 0.47). Additionally, some of these dissolved teams were formed by instructors and assigned as 'lone wolves' due to their inability to find partners in time. As a result, three out of the six randomly created teams dropped out. When students were asked about this situation, they indicated potential communication barriers arising from language differences, lack of previous interactions, and challenges in coordinating schedules.

On the other hand, the presence of women in the course represents a percentage of 27~\%, higher than that reported by Germany’s Federal Statistical Office (Destatis), which indicates that women made up nearly 24.9~\% of students in science and engineering courses at German higher education institutions during the first semester of the 2023/24 academic year \cite{destatis2023students}. 

\begin{figure}[h] 
 	\centering 
 	\includegraphics[width=0.7\columnwidth]{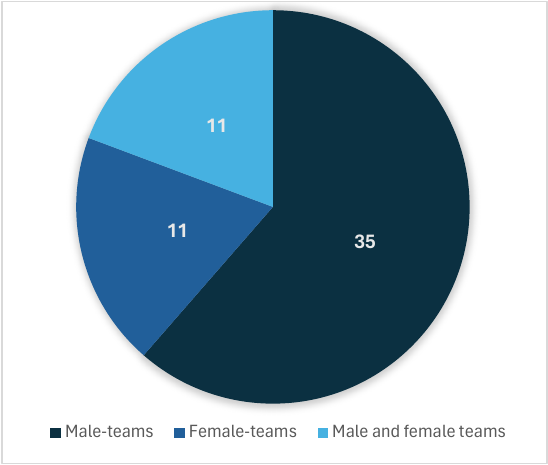} 
 	\caption{The distribution of the trend to form teams according to the members' gender, highlighting the prevalence of same-gender teams and the occurrence of mixed-gender teams within the student cohort, reflecting diverse team compositions based on gender dynamics.}
 	\label{gender} 
 \end{figure}

Another exciting pattern observed in team formation relates to the members' gender. Figure \ref{gender} illustrates that within a cohort of 33 women and 81 men, same-gender teams constituted 43, while mixed-gender teams comprised 11, representing approximately 19~\% of the total teams. When we asked about the selection of the team partner, the answer suggested that there was no specific reason. We did not find any significant difference in the obtained grades between these two types of teams (r = -0.12; p = 0.35). 
 
\subsection{Git Logs Analysis} 
 
Another interesting pattern identified is the distribution of commits made by students from the project release date to the submission deadline, spanning a period of seven weeks.

\begin{figure}[h] 
 	\centering 
 	\includegraphics[width=1\columnwidth]{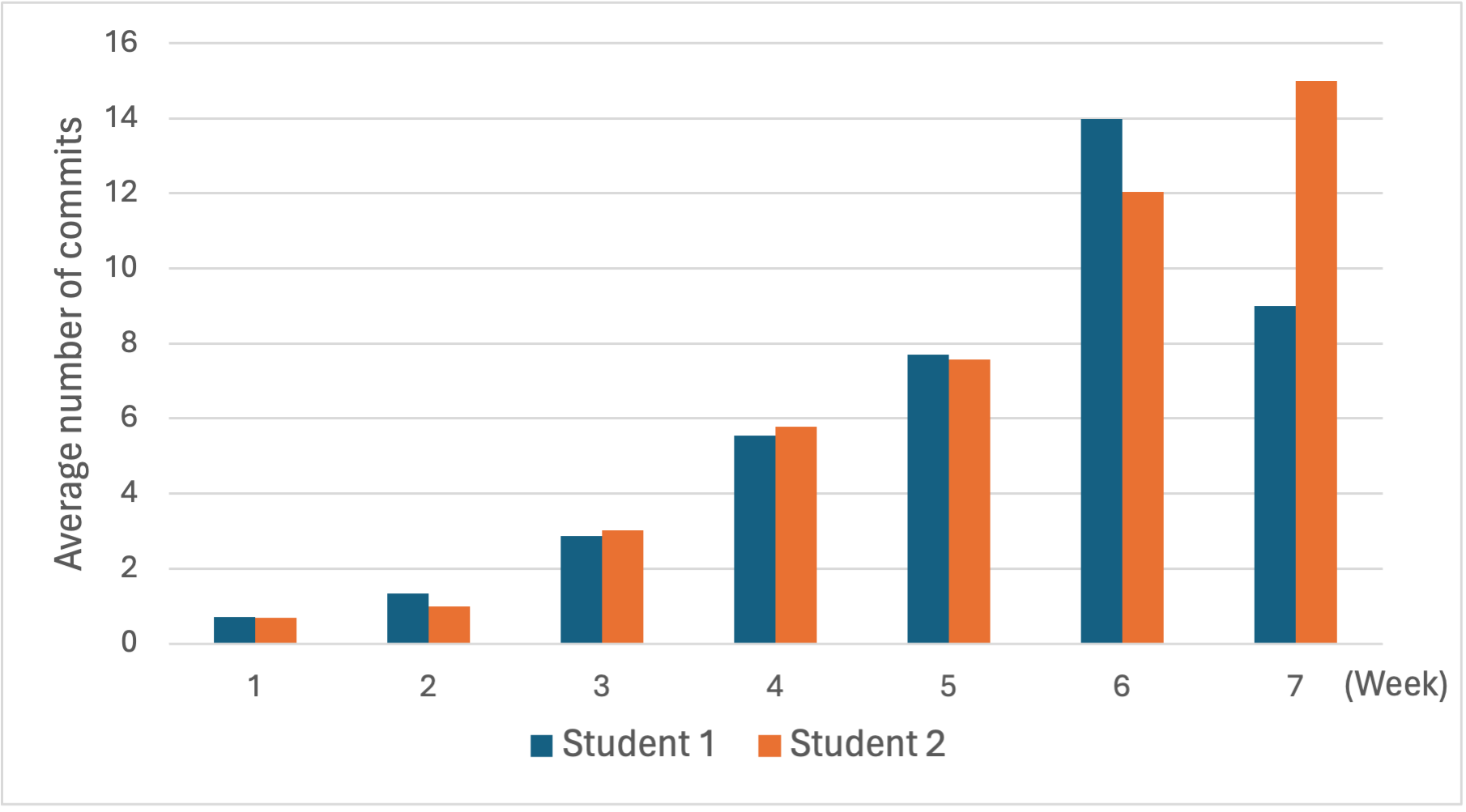} 
 	\caption{The average distribution of student commits throughout the project development period.}
 	\label{commitdistribution} 
 \end{figure}

As depicted in Figure \ref{commitdistribution}, the initial weeks show a sparse number of commits, which aligns with expectations. However, a noticeable trend has emerged in the last four weeks, revealing an almost linear increase in number of commits from each student. Moreover, during the first five weeks, there is minimal variance in collaboration among students. By week six, an average difference of two commits begins to emerge, which escalates to an average difference of six commits by week seven. This trend underscores progressive student engagement and commitment intensification as the project deadline approaches. 
 
 \subsection{Personal Contribution Estimation} 
 
\begin{figure}[h] 
 	\centering 
 	\includegraphics[width=1\columnwidth]{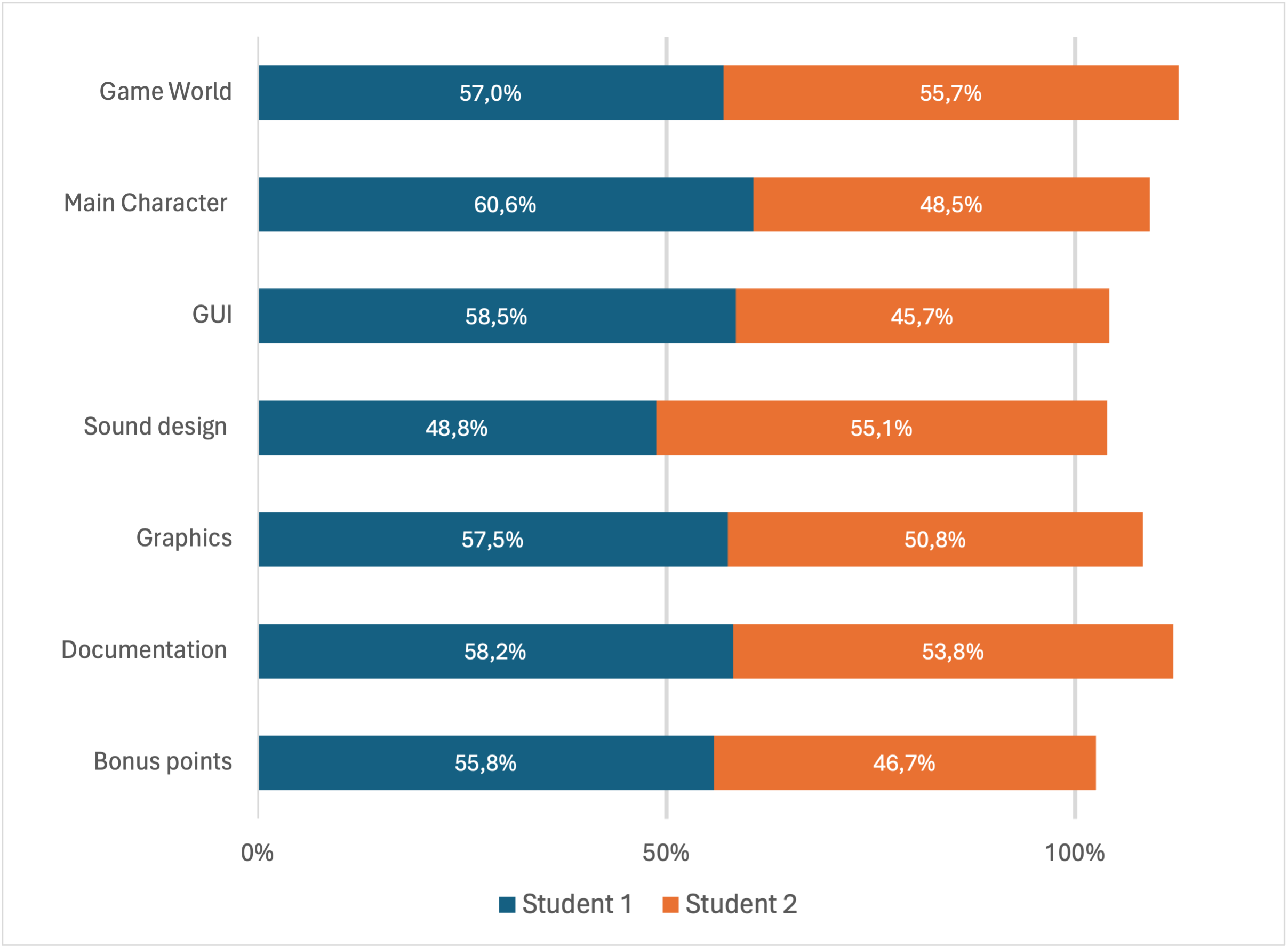} 
 	\caption{Estimation of students' self-assessment of their contributions to the project's components.}
 	\label{contribution} 
 \end{figure}

 Another relevant result was the students' estimation of their contributions to each project requirement. In Figure \ref{contribution}, it can be observed that the total contributions exceed 100~\% when aggregating individual estimates. Particularly noteworthy is the overestimation in the development of the Game World (112.7~\%) and the documentation (112~\%), indicating areas where contributions are most exaggerated. Conversely, the requirements for bonus points (102.5~\%), sound design (103.9~\%), and GUI (104.2~\%) are less overestimated.
 
\begin{figure}[h] 
 	\centering 
 	\includegraphics[width=1\columnwidth]{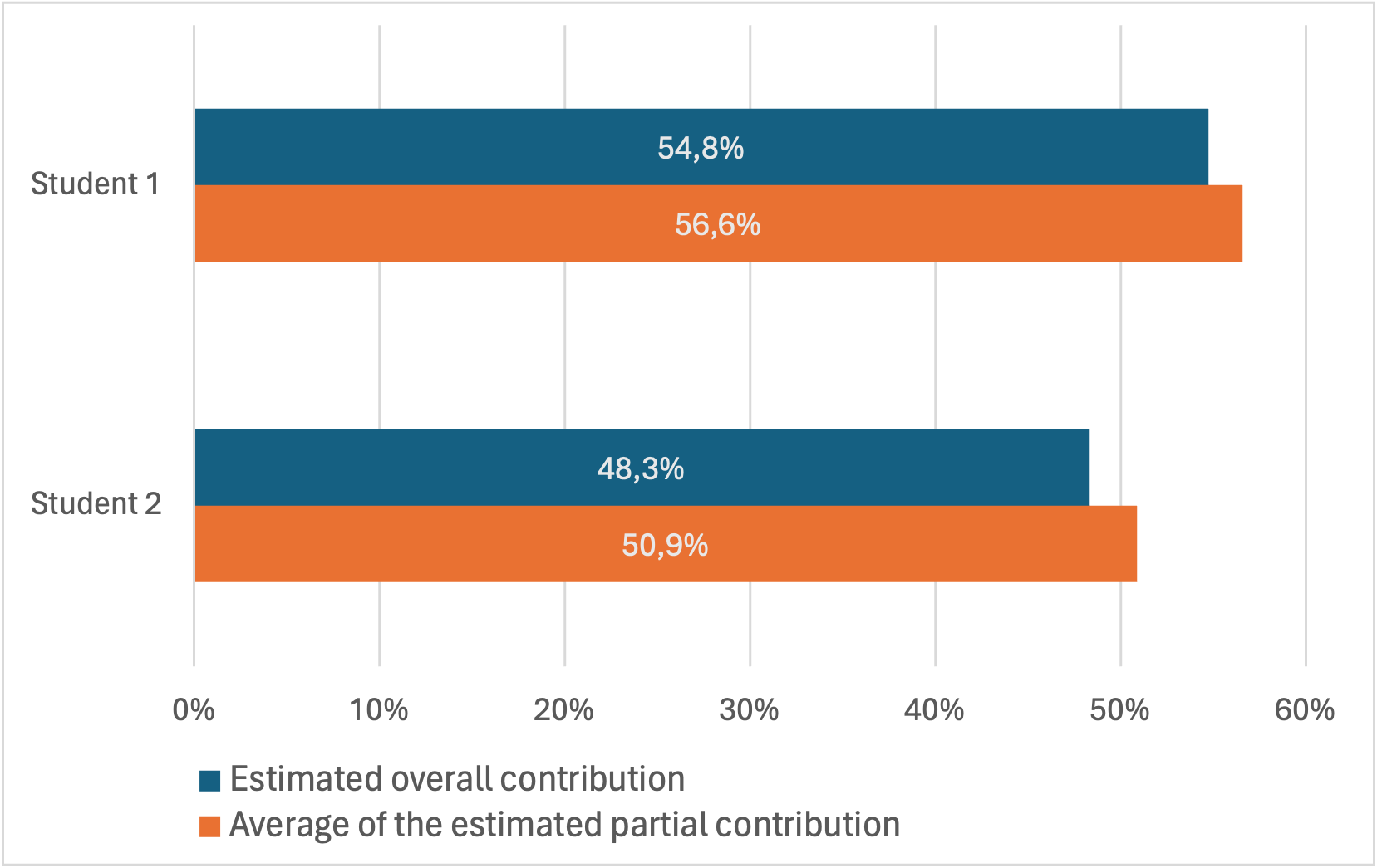} 
 	\caption{Comparison of the estimated overall contribution with the contribution obtained by summing up the partial estimated contributions.}
 	\label{overalcontribution} 
 \end{figure}

We found that when adding up each student's contributions to each requirement, the total differs from the overall contribution estimated by the same student. Figure \ref{overalcontribution} shows that, on average, each student's estimated contributions per requirement are overestimated compared to their total estimated contribution (1.8~\% -- 2.6~\%). 
 
\begin{figure}[h] 
	\centering 
	\includegraphics[width=1\columnwidth]{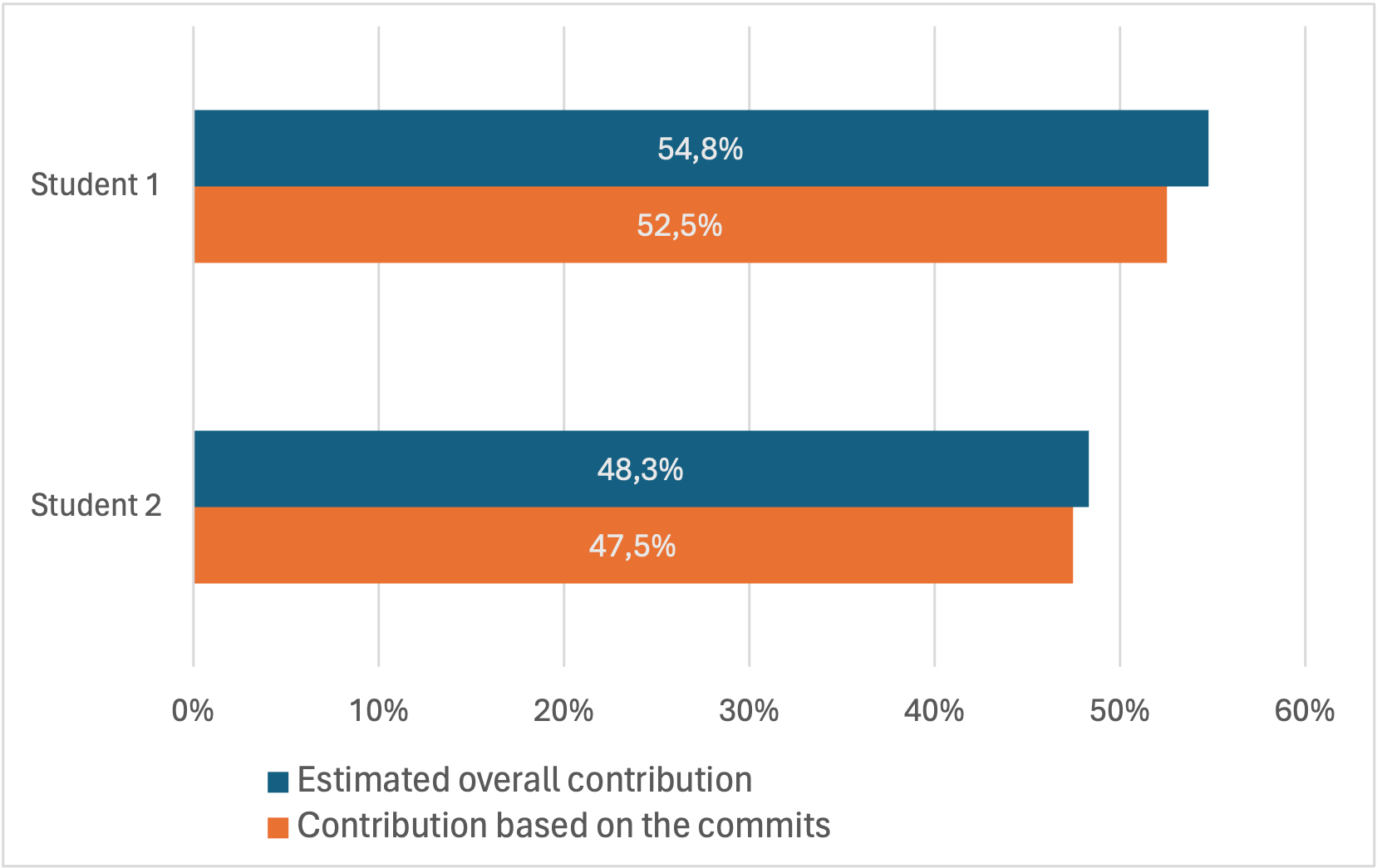} 
	\caption{Comparison of the estimated overall contribution with the real contribution obtained by analyzing the number of commits in the project.}
	\label{overalcontributioncommits} 
\end{figure}

We analyzed the commit history in each team repository to identify the most realistic estimation of each team member's contribution. We found notable trends by comparing the number of commits made by each student with their estimated overall contribution. Student 1 tended to overestimate their contribution by an average of 2.3 \%. Conversely, student 2 overestimated by an average of 0.8 \%. These findings are visualized in Figure \ref{overalcontributioncommits}.

These discrepancies suggest that students' perceptions of their contributions may sometimes align differently with their actual engagement, as reflected in their commit histories. Factors such as individual perceptions of effort, visibility of contributions, and interpretation of project responsibilities could contribute to these differences in estimation accuracy.
 
\begin{figure}[h] 
	\centering 
	\includegraphics[width=1\columnwidth]{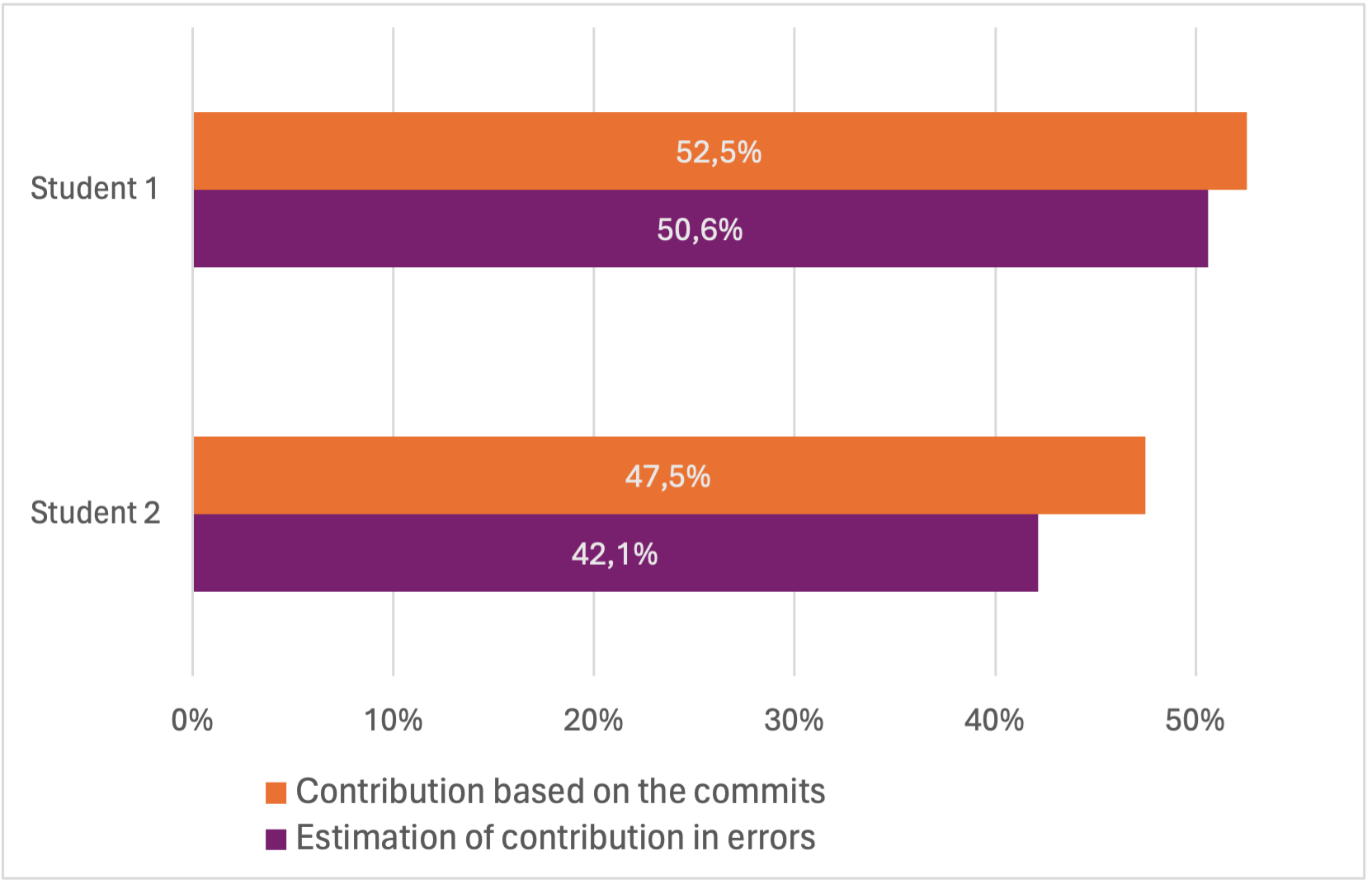} 
	\caption{Comparison between actual contributions, as determined by the number of commits in the project, and the students' estimated contributions to errors during project development.}
	\label{errors} 
\end{figure}

Figure \ref{errors} presents the estimation of errors made by students, which is not exaggerated. When combining the error contributions of both students, the average total is lower than 100 \%, approximately 92 \%. Unfortunately, there is no objective way to estimate these error contributions, as the backlog in each repository does not show a significant number of build failures. However, when comparing the error estimation with the contribution analysis from the backlog, we identified that students with more commits tend to estimate their fault contributions more significantly. In contrast, those with fewer contributions (commits) tend to underestimate their contributions to errors.

 \subsection{Use of Resources in the Teams} 
 
\begin{figure}[h] 
	\centering 
	\includegraphics[width=1\columnwidth]{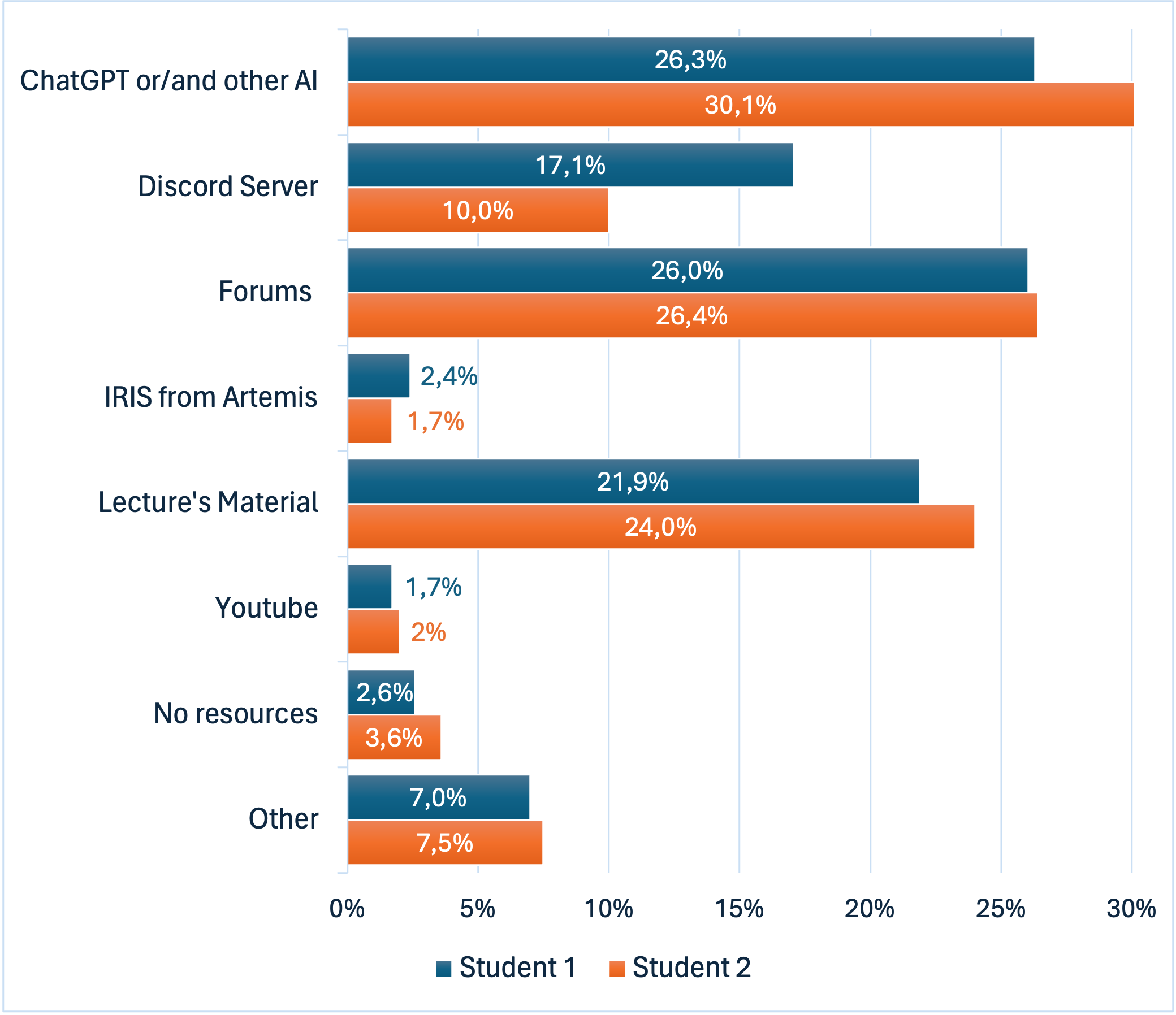} 
	\caption{Results of the analysis of the use of resources used by the members of the teams.}
	\label{resources} 
\end{figure}

 Figure \ref{resources} illustrates that ChatGPT and other AI tools were the preferred resources for project development. Close behind, forums such as Stack Overflow and lecture materials were also frequently utilized. The dedicated Discord server also played a crucial role in resolving queries, although only one team member typically posed the questions.
 
 It was noted that some students did not utilize any resources, while others opted for alternative methods such as consulting books and peers or directly engaging with instructors.
 
 Interestingly, YouTube and the IRIS\footnote{A chat-based virtual tutor integrated into the interactive learning platform Artemis that offers personalized, context-aware assistance in large scale educational settings \cite{bassner2024iris}.} of Artemis were notably less used, indicating a shift away from these resources in the project's context. 

\subsection{Soft Skills Evaluation}

During the project presentations, we observed a correlation where students who spoke the most during their presentations tended to estimate their contributions more highly in the subsequent survey (Perception contribution analysis). This phenomenon suggests that active participation during the presentation may lead individuals to perceive their involvement as more significant than it objectively may be. 
Additionally, the findings revealed that 76~\% of students with more significant commits were more effective during the question-and-answer (Q\&A) session. This suggests that students who contributed substantially to the project, as evidenced by their number of commits, were better prepared to discuss and defend their work during the presentation. 

 \subsection{Plagiarism Analysis}
 
We run the plagiarism check functionality of Artemis to identify any misconduct in developing the project. However, the plagiarism analysis among the teams revealed that the highest similarity score was below 37~\% similarity. Given the project's scope and the use of a standardized template with uniform requirements across all teams, these instances did not meet the criteria for plagiarism. This result underscores the effectiveness of informing students about the plagiarism check before the project began, which set the expectation for each team to produce independent work \cite{berrezueta2023plagiarism}.
 
\subsection{Individual Examination Analysis}
 
When analyzing the grades obtained in the project and the two individual intermediate exams, notable patterns emerged: among students who failed or did not complete one or both intermediate exams, 48~\% also failed the project, while 32~\% passed the project but with deficient grades, with none achieving a perfect score.

Conversely, students who scored above 80~\% in the intermediate exams showed a strong correlation with similar scores in the project, with 70~\% achieving comparable grades in both assessments.

Furthermore, examining the relationship between project grades and final exam outcomes revealed that 70~\% of students who scored above 80~\% in the project also passed the final exam. The calculated odds ratio of approximately 2.33 suggests that students achieving high scores in their project were about 2.33 times more likely to pass the final exam compared to those with lower project scores, assuming a 50~\% pass rate for the latter group. This finding underscores the significant positive association between high project performance and subsequent success in final exams.

\begin{figure}[h] 
	\centering 
	\includegraphics[width=0.8\columnwidth]{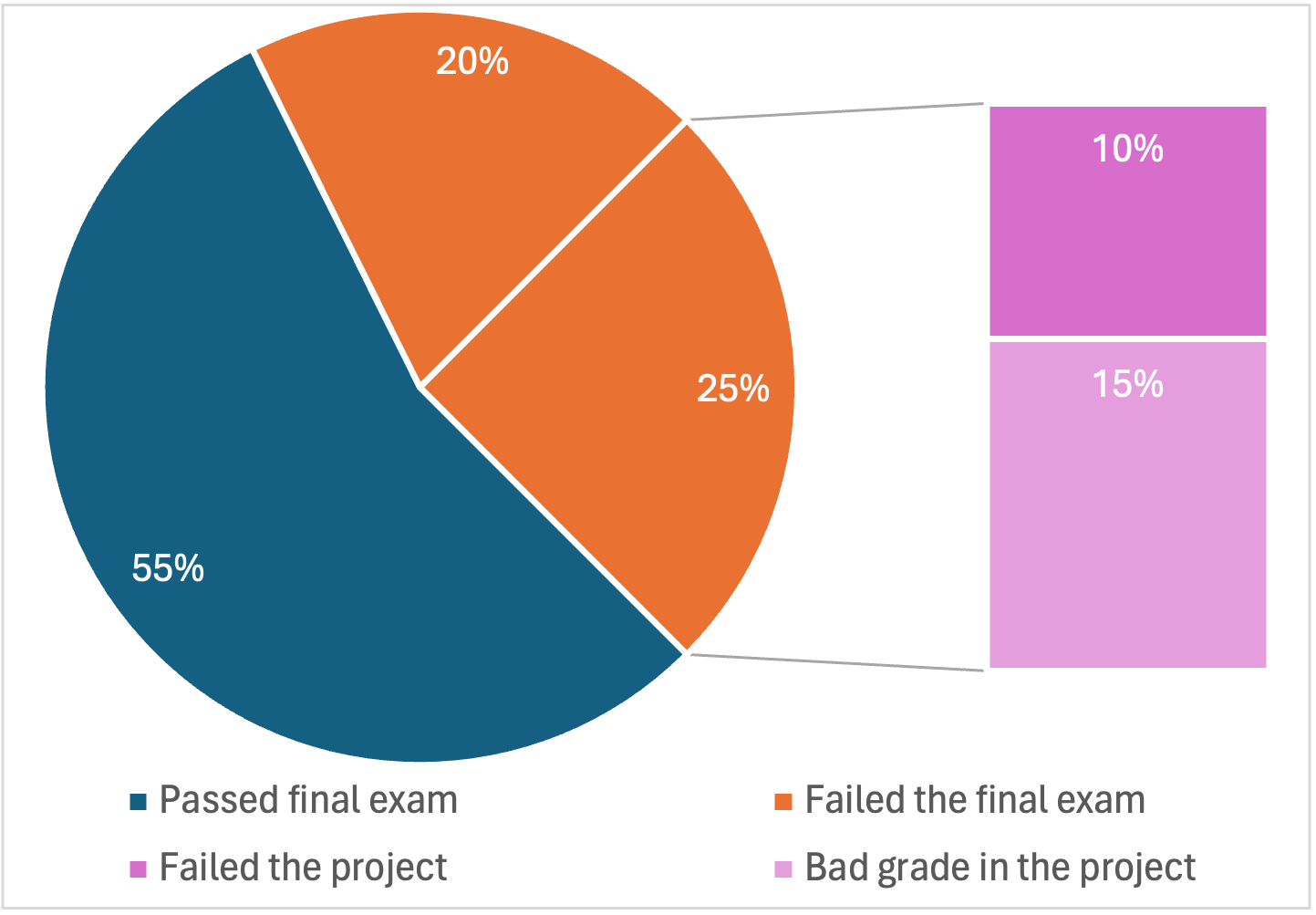} 
	\caption{Results of the analysis rating pass of the final exam with the rating pass of the project.}
	\label{finalExam} 
\end{figure}

Finally, we can observe in Figure \ref{finalExam} an important pattern regarding passing a project and a final exam. Around 55~\% of the students passed the final exam, and none of them failed the project. In contrast, within the 45~\% who failed the final exam, 20~\% passed the project. Additionally, 10~\% of the students who failed the exam failed the project. Finally, 25 \% of the students who failed the exam passed the project but received a grade under the average. 

Therefore, a moderate positive correlation exists between passing the final exam and performing well on the project of 0.337. Students who passed the final exam tended to perform well on the project, while those who failed the final exam were more likely to encounter difficulties with the project.

\section{Discussion}\label{D}

This section delves into key findings of the study on team dynamics and individual contributions to educational projects.

\note{\textbf{Finding 1:} There is no significant correlation between teams' performance based solely on nationality or gender. However, teams formed by ’lone wolves’ tended to quit the project.}

Interestingly, the teams that completely abandoned the project often consisted of 'lone wolves' from diverse nationalities. This highlights challenges associated with randomly assigned partnerships and underscores the need for individualized support to enhance collaborative skills among these students for future team projects. This analysis directly addresses RQ1 and RQ2 and supports previous research where no significant differences were found in the performance of software teams based on nationality or gender \cite{fernandez2012analysis}. 

\note{\textbf{Finding 2:} Students tend to overestimate their partial contributions, but their estimates of total contribution to the project are less exaggerated, as illustrated in Figure \ref{overalcontribution} and Figure \ref{commitdistribution}.}

However, in both cases, the overestimation is not extreme, averaging between 2.5~\% and 12.5~\% above the actual contributions, thereby addressing RQ3. Previous research also supports these results, which called this trend as \textit{egocentric bias} \cite{ramin2020role}. 

\note{\textbf{Finding 3:} Students who contribute more significantly to the project also tend to attribute to themselves the responsibility for the errors during development.}

This was not the case with students whose contributions were minimal, addressing RQ4. 
These overestimations could be influenced by students' eagerness to highlight their involvement, a lack of clear delineation between individual and team contributions, or a tendency to perceive their efforts more positively than objectively accurately. These findings underscore the importance of aligning students' perceptions with objective measures to better assess their contributions to collaborative projects.

\note{\textbf{Finding 4:} The decreasing use of platforms like YouTube among programming students suggests a shift in their preferences or a decline in the effectiveness of these resources for academic purposes in this particular course.}

This change also may be due to the project's specific requirements or the perceived immediacy and relevance of the content offered by these platforms. Until last year, programmers found it easier to follow a video than a textual post that might contain more jargon \cite{arya2023programmers}. However, now we can see that new first-year coders opt mainly to use ChatGPT or any other AI copilot when they want to program something. 
Such trends prompt further inquiry into how educational resources evolve and how student preferences shape the learning environment, addressing RQ5 and supporting previous research where the authors discovered that when students use ChatGPT, their inclination to explore other traditional educational resources significantly decreases. However, they manifested that this reliance on ChatGPT did not result in improved learning performance \cite{xue2024does}.

\note{\textbf{Finding 5:} Students who contributed more substantially to the development of the project could present more effectively and respond confidently to evaluators' questions.}

The results from the project presentation scores show a strong correlation with each team member's contributions. The focus was identifying students who did not significantly contribute to the project (based on the Git Logs analysis) and directing questions toward them. As expected, these students needed help answering questions from the evaluation panel, which addressed RQ6.

\note{\textbf{Finding 6:} The absence of plagiarism cases confirms the effectiveness of the preventative measures outlined in \cite{berrezueta2023plagiarism}.}

These measures include in-advance warnings about similarity checks for all teams at the end of the deadline and clear communication of consequences. Therefore, the students avoided misconduct by sharing their code. This finding addressed RQ7.

\note{\textbf{Finding 7:} Students with significant contributions to the project were likely to pass their examinations. In contrast, students with minimal contributions failed the exams and, consequently, the course.}

This list also includes students who abandoned the project, suggesting that team projects can predict individual examination outcomes, addressing research question 8. 

The \textbf{Limitations} of this study include a relatively small sample size of 57 groups with a limited number of female students. Additionally, the study's focus on a single institution may limit the applicability of the results, as educational practices and student behavior can vary across different regions and institutions. The project duration was brief, lasting only seven weeks, which might not fully capture the long-term dynamics and development of teamwork skills. Therefore, the findings, while indicative, should not be generalized to the entire population of first-semester students. Although the sample was diverse in nationality, it was homogeneous in academic background, with most students having prior experience in computer science. This homogeneity may limit the findings' relevance to students from other academic disciplines. 

\section{Conclusion}\label{C}

This study highlights the critical role that collaborative projects play in introductory computer science education, offering valuable insights into team dynamics, individual contributions, and their impact on academic outcomes. Our findings reveal that team composition, based on factors such as nationality and gender, does not significantly affect project performance. Instead, student success is influenced more by engagement levels, task management, and interpersonal dynamics within teams, regardless of demographic factors. This underscores the value of fostering inclusive teamwork environments that encourage equal participation.

The alignment between active participation, project presentations, and exam performance underscores the importance of practical engagement in strengthening understanding and communication skills. Additionally, the absence of plagiarism validates the effectiveness of preventative integrity measures, reinforcing the need for transparency in academic evaluation.

The positive correlation between project engagement and final exam success highlights the pedagogical potential of collaborative projects as predictors of individual academic achievement. These findings advocate for integrating project-based learning into computer science curricula to foster technical skills and develop critical soft skills like teamwork, problem-solving, and self-assessment. Our study suggests that well-structured collaborative projects can be powerful tools in developing students’ technical and interpersonal skills, better preparing them for professional environments. 

\textbf{Future work} should examine how teamwork dynamics change when students work in larger groups, such as teams of five or more, compared to the pairs used in this study. Increasing team size introduces new variables, such as task distribution complexity, conflict resolution, and group coordination, which may affect project outcomes and individual engagement. Additionally, a longitudinal study tracking students’ teamwork skills across multiple semesters would be valuable. Following students through their academic journey could reveal patterns in skill acquisition and highlight the long-term benefits of structured collaborative learning.

\bibliographystyle{ieeetr}
\bibliography{PaperTeamCollaboration}

\begin{thebibliography}{10}

\bibitem{oakley2004turning}
B.~Oakley, R.~M. Felder, R.~Brent, and I.~Elhajj, ``Turning student groups into
  effective teams,'' {\em Journal of student centered learning}, vol.~2, no.~1,
  pp.~9--34, 2004.

\bibitem{chorfi2022problem}
A.~Chorfi, D.~Hedjazi, S.~Aouag, and D.~Boubiche, ``Problem-based collaborative
  learning groupware to improve computer programming skills,'' {\em Behaviour
  \& Information Technology}, vol.~41, no.~1, pp.~139--158, 2022.

\bibitem{prada2022teamwork}
E.~D. Prada, M.~Mareque, and M.~Pino-Juste, ``Teamwork skills in higher
  education: is university training contributing to their mastery?,'' {\em
  Psicologia: Reflexao e critica}, vol.~35, p.~5, 2022.

\bibitem{yadav2021collaborative}
A.~Yadav, C.~Mayfield, S.~K. Moudgalya, C.~Kussmaul, and H.~H. Hu,
  ``Collaborative learning, self-efficacy, and student performance in cs1
  pogil,'' in {\em Proceedings of the 52nd ACM Technical Symposium on Computer
  Science Education}, pp.~775--781, 2021.

\bibitem{soundarajan2015collaborative}
N.~Soundarajan, S.~Joshi, and R.~Ramnath, ``Collaborative and
  cooperative-learning in software engineering courses,'' in {\em 2015 IEEE/ACM
  37th IEEE International Conference on Software Engineering}, vol.~2,
  pp.~319--322, IEEE, 2015.

\bibitem{neill2017improving}
C.~J. Neill, J.~F. DeFranco, and R.~S. Sangwan, ``Improving collaborative
  learning in online software engineering education,'' {\em European Journal of
  Engineering Education}, vol.~42, no.~6, pp.~591--602, 2017.

\bibitem{gitinabard2020student}
N.~Gitinabard, R.~Okoilu, Y.~Xu, S.~Heckman, T.~Barnes, and C.~Lynch, ``Student
  teamwork on programming projects: What can github logs show us?,'' {\em arXiv
  preprint arXiv:2008.11262}, 2020.

\bibitem{gitinabard2023analysis}
N.~Gitinabard, Z.~Gao, S.~Heckman, T.~Barnes, and C.~F. Lynch, ``Analysis of
  student pair teamwork using github activities.,'' {\em Journal of Educational
  Data Mining}, vol.~15, no.~1, pp.~32--62, 2023.

\bibitem{sancho2009learning}
P.~Sancho-Thomas, R.~Fuentes-Fern{\'a}ndez, and B.~Fern{\'a}ndez-Manj{\'o}n,
  ``Learning teamwork skills in university programming courses,'' {\em
  Computers \& Education}, vol.~53, no.~2, pp.~517--531, 2009.

\bibitem{richards2012cross}
D.~Richards and A.~Bilgin, ``Cross-cultural study into ict student attitudes
  and behaviours concerning teams and project work,'' {\em Multicultural
  Education \& Technology Journal}, vol.~6, no.~1, pp.~18--35, 2012.

\bibitem{iacob2019exploring}
C.~Iacob and S.~Faily, ``Exploring the gap between the student expectations and
  the reality of teamwork in undergraduate software engineering group
  projects,'' {\em Journal of systems and software}, vol.~157, p.~110393, 2019.

\bibitem{simpson2017experimenting}
R.~Simpson and T.~Storer, ``Experimenting with realism in software engineering
  team projects: An experience report,'' in {\em 2017 IEEE 30th Conference on
  Software Engineering Education and Training (CSEE\&T)}, pp.~87--96, IEEE,
  2017.

\bibitem{krusche2017interactive}
S.~Krusche, A.~Seitz, J.~B{\"{o}}rstler, and B.~Br{\"{u}}gge, ``Interactive
  learning: Increasing student participation through shorter exercise cycles,''
  in {\em Proceedings of the 19th Australasian Computing Education Conference},
  pp.~17--26, {ACM}, 2017.

\bibitem{krusche2023introduction}
S.~Krusche and J.~Berrezueta-Guzman, ``Introduction to programming using
  interactive learning,'' in {\em 2023 IEEE 35th International Conference on
  Software Engineering Education and Training (CSEE\&T)}, pp.~178--182, IEEE,
  2023.

\bibitem{krusche2018artemis}
S.~Krusche and A.~Seitz, ``Artemis: An automatic assessment management system
  for interactive learning,'' in {\em Proceedings of the 49th ACM technical
  symposium on computer science education}, pp.~284--289, 2018.

\bibitem{linhuber2024}
M.~Linhuber, J.~P. Bernius, and S.~Krusche, ``Constructive {{Alignment}} in
  {{Modern Computing Education}}: {{An Open-Source Computer-Based Examination
  System}},'' in {\em Proceedings of the 23rd {{Koli Calling International
  Conference}} on {{Computing Education Research}}}, Koli {{Calling}} '23, ACM,
  2024.

\bibitem{prechelt2000jplag}
L.~Prechelt, G.~Malpohl, and M.~Philippsen, ``Jplag: Finding plagiarisms among
  a set of programs,'' 2000.

\bibitem{stemkoski2018libgdx}
L.~Stemkoski and L.~Stemkoski, ``The libgdx framework,'' {\em Java Game
  Development with LibGDX: From Beginner to Professional}, pp.~15--37, 2018.

\bibitem{berrezueta2023recommendations}
J.~Berrezueta-Guzman and S.~Krusche, ``Recommendations to create programming
  exercises to overcome chatgpt,'' in {\em 2023 IEEE 35th International
  Conference on Software Engineering Education and Training (CSEE\&T)},
  pp.~147--151, IEEE, 2023.

\bibitem{destatis2023students}
S.~Bundesamt, ``Students enrolled in stem courses,'' Aug 2023.
\newblock Accessed: 2024-05-16.

\bibitem{bassner2024iris}
P.~Bassner, E.~Frankford, and S.~Krusche, ``Iris: An ai-driven virtual tutor
  for computer science education,'' {\em arXiv preprint arXiv:2405.08008},
  2024.

\bibitem{berrezueta2023plagiarism}
J.~Berrezueta-Guzman, M.~Paulsen, and S.~Krusche, ``Plagiarism detection and
  its effect on the learning outcomes,'' in {\em 2023 IEEE 35th International
  Conference on Software Engineering Education and Training (CSEE\&T)},
  pp.~99--108, IEEE, 2023.

\bibitem{fernandez2012analysis}
L.~Fernandez-Sanz and S.~Misra, ``Analysis of cultural and gender influences on
  teamwork performance for software requirements analysis in multinational
  environments,'' {\em IET software}, vol.~6, no.~3, pp.~167--175, 2012.

\bibitem{ramin2020role}
F.~Ramin, ``The role of egocentric bias in undergraduate agile software
  development teams,'' in {\em Proceedings of the ACM/IEEE 42nd International
  Conference on Software Engineering: Companion Proceedings}, pp.~122--124,
  2020.

\bibitem{arya2023programmers}
D.~M. Arya, J.~L. Guo, and M.~P. Robillard, ``How programmers find online
  learning resources,'' {\em Empirical Software Engineering}, vol.~28, no.~2,
  p.~23, 2023.

\bibitem{xue2024does}
Y.~Xue, H.~Chen, G.~R. Bai, R.~Tairas, and Y.~Huang, ``Does chatgpt help with
  introductory programming? an experiment of students using chatgpt in cs1,''
  in {\em Proceedings of the 46th International Conference on Software
  Engineering: Software Engineering Education and Training}, pp.~331--341,
  2024.

\end{thebibliography}
\end{document}